\documentstyle[preprint,aps]{revtex}
\begin{document}
\draft
\preprint{}
\title{Tunneling into a two-dimensional electron system in a strong
magnetic field }
\author{Song~He, P.~M.~Platzman}
\address{
AT\&T Bell Laboratories, Murray Hill, NJ 07974 \\
}
\author{B.~I.~Halperin}
\address{
Department of Physics, Harvard University \\
Cambridge, MA 02138
}
\date{April 13, 1993}
\maketitle
\begin{abstract}
We investigate the properties of the one-electron Green's function in an
interacting  two-dimensional electron system in a strong
magnetic field, which describes an
electron tunneling into such a system. From finite-size diagonalization, we
find
that its spectral weight is suppressed near zero energy, reaches a maximum at
an energy of about $0.2e^{2}/\epsilon l_{c}$, and decays exponentially at
higher energies.
We propose a theoretical model to account for the low-energy behavior. For the
case of
Coulomb interactions between the electrons, at even-denominator filling
factors such as $\nu=1/2$, we predict that
the spectral weight varies as $e^{-\omega_0/|\omega|}$, for $\omega\rightarrow
0$.
\end{abstract}

\narrowtext
High mobility $e^+$ or $e^-$ doped GaAs-GaAlAs quantum wells constitute a
remarkable, almost ideal,
many-body system. In such a system, Coulomb interactions among the electrons
play a vital role in determining its properties. This is particularly true when
the system is immersed in a strong perpendicular magnetic field which quenches
the kinetic energy of the electrons, the so-called fractional quantum Hall
regime. In order to experimentally probe the properties of such systems,
investigations have primarily utilized measurements of the magnetotransport
coefficients\cite{FQHE}.
However, within the last year, a variety of spectroscopic probes have begun to
play an important role\cite{Aron}\cite{Junk}\cite{Ray}\cite{Jim0}.

A recent experiment \cite{Jim0} measured the low temperature high field bulk
tunneling characteristics of a bilayer system with electron density
$n\equiv\nu/2\pi l_c^2$ in each layer, coupled together by a weak tunnel
barrier, separated by a distance $d\approx 2n^{-1/2}$. The experimental I-V
characteristics in the range $0.48<\nu<0.83$ exhibited a strong suppression of
the tunneling current at low biases, similar to what was observed in Ref.
\cite{Ray},
and a broad, pronounced peak in the
neighborhood of $eV_{max}\sim0.4e^2/\epsilon l_c$. Here $\epsilon$ is the
dielectric constant, and $l_c$ is the magnetic length.

In this paper we construct a theory of this type of tunneling experiment. In
particular, we present exact numerical results on the one-electron Green's
function for a small number of particles which we believe gives reliable
information at energies comparable to $eV_{max}$. At lower energies we
construct a
model which directly relates the I-V curve to the low-lying density
fluctuations in the system. A comparison of the theory with experiment is made.

Since the layer separation $d$ is large, we can, to the first approximation,
ignore coupling between the layers. In this limit, it may be shown that for
electrons confined to the lowest Landau level, the tunneling current obtained
from the general Golden rule expression becomes:
\begin{equation}
\label{IV}
I=e\lambda^2\Omega l_c^{-2}\int_0^{eV} d\omega A_+(\omega) A_-(eV-\omega),
\end{equation}
where $V$ is the voltage difference between the layers, $\lambda$ is the
tunneling matrix element, $\Omega$ is the area of the system, and
$A_{\pm}(\omega)$ are the spectral weights to add (substract) an electron to
(from) the right (left) layer. Written in terms of the operators
$\hat{\psi}^\dagger$ and $\hat{\psi}$ which creates or annihilates an electron
at the origin, these spectral weights are:
\begin{eqnarray}
A_{+}(\omega)&=&
\sum_{\alpha}|\langle\alpha,N+1|\hat{\psi}^{\dagger}|0,N\rangle|^2\nonumber\\
&\times&\delta(\omega+\mu+E_{0,N}-E_{\alpha,N+1}),
\end{eqnarray}
\begin{eqnarray}
A_{-}(\omega)&=&
\sum_{\alpha}|\langle\alpha,N-1|\hat{\psi}|0,N\rangle|^2\nonumber\\
&\times&\delta(\omega-\mu+E_{0,N}-E_{\alpha,N-1}),
\end{eqnarray}
where $\mu\equiv(\mu_++\mu_-)/2$, $\pm\mu_{\pm}=E_{0,N\pm1}-E_{0,N}$, and
$\mu_{\pm}$ are the chemical potentials for adding and subtracting an electron
form an $N$-electron system respectively. Note that for an incompressible
state, the chemical potential jump $\Delta_\mu\equiv\mu_+-\mu_->0$. Therefore
the low-bias suppression of the tunneling current observed in the experiment
implies that the low-energy spectral weight of the one-electron Green's
function must vanish strongly for $\omega\rightarrow 0$, in contrast to the
behavior of a normal Fermi liquid. We shall see that this behavior can be
understood using rather simple arguments.

The mean value $\bar{\omega}_+$ of the energy in $A_+(\omega)$ is determined by
the expectation value of the Hamiltonian $\hat{H}$ in the ``initial state'',
$|\Phi_1\rangle\equiv\hat{\psi}^\dagger|0,N\rangle$. We may estimate
this if we consider $|\Phi_1\rangle$ to be an assemblage of quasiparticles,
packed together as
closely as possible to give one extra electron charge in a disc about the
origin. Since the maximum electron density in any Landau level is $(2\pi
l_c^2)^{-1}$, and the background charge
is $\nu (2\pi l_c^2)^{-1}$, the minimum radius of the charged disc is
$r_+=l_c\sqrt{2/(1-\nu)}$. We may estimate $\bar{\omega}_+$ as being equal to
the Coulomb self
energy of this disc, or roughly $\bar{\omega}_+\approx 0.2e^2/l_c$ for
$\nu\approx 1/2$. The mean energy $\bar{\omega}_-$ of $A_-(\omega)$ may be
estimated similarly assuming the hole charge is spread out in a disc of radius
$r_-=l_c\sqrt{2/\nu}$. If the spectral weights are
concentrated near $\bar{\omega}_{\pm}$, then the tunneling current should have
a maximum at $eV_{max}\approx \bar{\omega}_+ + \bar{\omega}_-\approx
0.4e^2/\epsilon l_c$, the coefficient being roughly
independent of $\nu$ in the range of $0.2<\nu<0.8$.

To understand crudely the suppression of $A_{\pm}(\omega)$ at small values of
$\omega$, consider a circular droplet of electron liquid with a uniform density
corresponding to filling factor $\nu$. In the symmetric gauge for the magnetic
vector potential, and in the absence of disorder, the total angular momentum
$M$ perpendicular to the layer is a conserved quantity. The value of $M$ of a
uniform droplet containing $N$ electrons is $M_{0,N}\approx N^2/2\nu$, so that
$M_{0,N+1}-M_{0,N}\approx N/\nu$. The initial state $|\Phi_1\rangle$ with an
electron added to the origin of the $N$-electron droplet has angular momentum
$M_{0,N}$, which is very different from $M_{0,N+1}$. Thus not only is the state
$|\Phi_1\rangle$ orthogonal to the groundstate of the $(N+1)$-electron droplet,
it should also have little overlap with any of the low-lying excited states. In
order for the low-energy excitations to carry away the necessary angular
momentum, they must be created far from the origin, which means that there is
little overlap with the initial state that is only perturbed at the origin.

If the filling fraction $\nu$ corresponds to a quantized Hall state with
denominator $2p+1$, then the elementary charged excitations are quasiparticles
with charge $e/(2p+1)$. Ignoring boundary excitations, and assuming a repulsive
interaction between the quasiparticles, we
expect that the lowest states of a droplet with one extra electron charge
contain $2p+1$ quasiparticles well separated from each other. In order to
obtain the correct total angular momentum, there must be at least one neutral
excitation, such as a magnetoroton\cite{SMA} in
addition. Thus, in the quantized Hall case there should be a true energy gap in
$A_+(\omega)$, with a threshold $\omega_+$ that is slightly larger than
$(2p+1)\tilde{\varepsilon}_+$, where $\tilde{\varepsilon}_+$ is the energy to
add a single quasiparticle. In Ref.\cite{HLR}, it was
proposed that for the principal quantized Hall states at $\nu=p/(2p+1)$, the
combination
$(2p+1)\tilde{\varepsilon}_+$ should be only weakly dependent on $p$, and the
threshold $\omega_+$ obtained in this way is only slightly smaller than the
mean energy $\bar{\omega}_{+}$ estimated above. For other quantized Hall states
the value of $\omega_+$ should be much smaller than this, and for a
compressible state the true threshold must occur at
$\omega=0$. Nevertheless, to obtain a low energy state with the correct total
charge and angular momentum, we must produce excitations far from the origin,
so that $A_+(\omega)$ should
be very small for $\omega\rightarrow 0$, as we find in our quantitive estimates
below.

In order to study $A_{\pm}(\omega)$ for energies around $\bar{\omega}_\pm$, we
have performed exact calculations of systems of up to 10 electrons in the
lowest Landau level in a spherical geometry. In the spherical geometry, due to
the rotational symmetry, the eigenstates
of the system can be classified by their total angular momentum $\hat{L}^2$ and
$\hat{L}_z$. We consider only systems where the groundstate of the initial $N$
particle system has a zero total angular momentum. Since an electron has an
intrinsic orbital angular momentum $l=S\equiv N_{\phi}/2$ when it is confined
to the
surface of a sphere with a magnetic monopole of total flux $N_{\phi}$ at the
center, we see that when
one more electron is added, the resultant $(N+1)$-electron system has a total
angular momentum $S$. Therefore the spectral weight is
nonzero only for states with angular momentum $S$.

In Fig.\ \ref{FIG1}, we show the histograms of the spectral weight $A(\omega)$,
defined as $A_+(\omega)$ for $\omega>0$ and $A_-(-\omega)$ for $\omega<0$, for
a compressible system of 8 electrons and at an $N_{\phi}=17$ close to
$\nu=1/2$. Apparently there are a few dominant peaks in $A_{\pm}(\omega)$ at
energies comparable to $0.2e^{2}/\epsilon l_{c}$. There
is no weight at energies below these peaks due to finite size effect: there are
simply no states with the correct angular momentum in this energy interval for
such a small system. The high energy behavior is shown in the inset. We find
$A(\omega)\sim e^{-|\omega|/\Gamma}$ with $\Gamma\approx 0.026e^2/\epsilon
l_c$. We attribute the high energy tail to states where the added electron is
accompanied by one or more
excitations of a short wavelength density mode within the lowest Landau level.
A calculation including higher Landau levels would presumably show additional
high-energy side bands shifted up by multiples of the cyclotron frequency. In
Fig.\ \ref{FIG2}, we show $A(\omega)$ for a system of 8 electrons at a filling
factor $\nu=1/3$. Note that in the case, the tunneling peak position is
$eV_{max}\approx 0.6e^2/\epsilon l_c$. We have also done calculations at
$\nu=1/3$ for systems with up to 10 electrons. We find by $1/N$ extrapolation
that $eV_{max}\approx 0.5e^2/\epsilon l_c$ for an infinite system at $\nu=1/3$.
Calculations at other filling factors in the range
$1/3\leq\nu\leq 2/3$ give results qualitatively similar to Fig.\ \ref{FIG1} and
Fig.\ \ref{FIG2}, and suggest a tunneling peak energy $eV_{max}\sim(0.4\sim
0.6)e^{2}/\epsilon l_{c}$ in all cases.

In order to predict the low-energy behavior in a compressible state, we employ
a number of approximations which reduce the physics of this problem to that of
the X-ray edge problem. We treat the added electron as if it were an infinitely
massive foreign particle inserted into the $N$-electron system. Physically this
should be a good approximation because of the large magnetic field which
suppresses the effect of exchange on a short distance scale and practically
eliminates recoil effects. If $\hat{H}_N$ is the Hamiltonian of the
$N$-electron system before insertion, the Hamiltonian
after the insertion has the form
\begin{equation}
\label{H}
\hat{H}=\hat{H}_N+E'+\sum_{q}V_q\hat{\rho}_q
\end{equation}
where $V_q=2\pi e^2/q$ is the Fourier transform of the Coulomb interaction, and
$\hat{\rho}_q$ is the density operator for the original $N$-electron system,
projected onto the lowest Landau level. The constant $E'$ is chosen so that the
groundstate energy of $\hat{H}$ coincides with the correct groundstate energy
of the $(N+1)$-electron system.

Following the method of Mahan\cite{Mahan}, Langreth\cite{Lan}, and Shung and
Langreth\cite{Lan},
treating the last term in Eq.(\ref{H}) as a weak perturbation in a linked
cluster expansion of the extra electron self-energy, we immediately arrive at
an analytic expression for the imaginary time electron Green's function
$G_+(\tau)=e^{-C(\tau)}$, where
\begin{equation}
\label{CC}
C(\tau)=\sum_q
V_q^2\int_{0}^{\infty}\frac{d\omega}{\pi}\frac{\text{Im}\chi}{\omega^2}(1-e^{-\tau\omega}).
\end{equation}
where $\chi(q,\omega)$ is the density response function of the electron system.
Since $G_+(\tau)$ is the Laplace transform of $A_+(\omega)$, Eq.(\ref{CC}) is
equivalent to replacing the electron system by a set of Harmonic oscillators
with the same density response\cite{Schotte}.

Since we are interested in the low-energy behavior of $A_+(\omega)$ ({\it i.e.}
long time behavior of $G_+(\tau)$), we need some estimate of $\chi(q,\omega)$
at low energies and long wavelengths. Recently
Halperin, Lee, and Read\cite{HLR} proposed a Chern-Simons Fermi liquid theory
of the $\nu=1/2$ state, which implies that, for Coulomb interaction, in the
absence of impurity scattering, $\chi(q,\omega)$ is dominated by a diffusive
mode at small $q$ and $\omega$, {\it i.e.}
\begin{equation}
\label{Chi}
\chi(q,\omega)=\frac{1}{V_q}\frac{1}{1-i\omega/\beta q^2},
\end{equation}
where, according to the RPA calculation in Ref.\cite{HLR},
$\beta=e^2l_c/4\epsilon$ for $\nu=1/2$. For a general compressible state, we
expect that Eq.(\ref{Chi}) still holds but with some modified value of $\beta$.
Substituting Eq.(\ref{Chi}) into Eq.(\ref{CC}), we obtain, for large $\tau$,
\begin{eqnarray}
\label{LT}
C(\tau)&\approx&\frac{e^2}{\pi\epsilon}\sqrt{\frac{\tau}{\beta}}\int_0^{\infty}dx\,dy \frac{1}{xy^2(1+x^2)}(1-e^{-xy^2})\nonumber\\
&\equiv&2\sqrt{\omega_0\tau},
\end{eqnarray}
where $\omega_0=\pi e^2/2\epsilon l_c$. Since $A_+(\omega)\geq 0$, the
$\tau\rightarrow\infty$ behavior of the Laplace transform $G_+$ determines the
low frequency behavior of $A_+$, and we find, for small $\omega$,
\begin{equation}
\label{AAA}
A_+(\omega)\sim e^{-\omega_0/\omega}.
\end{equation}
The identical result is obtained for the hole contribution, $A_-(\omega)$.
Combining this with Eq.(\ref{IV}), we find at small bias $V$, at $\nu=1/2$, the
tunneling current is
\begin{equation}
\label{RR}
I\sim\int_{0}^{eV}d\omega
\exp{[-\omega_0(\frac{1}{\omega}+\frac{1}{eV-\omega})]}\sim e^{-V_0/V}.
\end{equation}
where $eV_0=4\omega_0$. Obviously, the tunneling current $I$ at small $V$
vanishes faster than any power law in $V$.

For general filling fraction close to $\nu=1/2$, we expect that
Eqs.(\ref{LT})(\ref{AAA})(\ref{RR}) should apply for intermediate times or
energies, and that eventually there will be a crossover to some other behavior.
For a pure system at a high order quantized Hall state, we expect to find a
true gap at sufficiently low energies; for a compressible state at any
even-denominator fraction that can be described as a Chern-Simons Fermi liquid,
the ultimate $\omega\rightarrow 0$ behavior will be similar to Eq.(\ref{AAA}),
but with a value of $\omega_0$ that depends on the filling fraction.
When impurity scattering is taken into account, one finds that at long wave
lengths, the relaxation rate
$\beta q^2$, in the denominator of Eq.(\ref{Chi}), must be replaced by the
two-dimensional dielectric relaxation rate, $2\pi q \sigma_{xx}/\epsilon$,
where $\sigma_{xx}$ is the diagonal conductivity at $\nu=1/2$. This
substitution leads to a modification of the low-energy behavior to
$A(\omega)\sim \exp{(-x[\ln{(2\pi\sigma_{xx}/\epsilon l_c |\omega|)}]^2)}$,
where $x\equiv e^2/(2\pi^2\sigma_{xx})$ is
of the order ten in typical experiments.

The interaction between layers may also be taken into account, although the
analysis is more complicated in this case. It appears that the most important
effect is a downward shift of the peak voltage $V_{max}$, by an amount
$e^2/\epsilon d$, which reflects a reduction of the mean energy of the initial
tunneling state due to the attraction between the electron and the hole.
In order to relax towards a uniform groundstate at large value of imaginary
time $\tau$, the electron and the hole must spread in space so that on average
they are far apart. The long time relaxation must be controlled by the
relaxation of the density difference in the two layers, which for a system
without impurities, and wavevectors $q \ll 1/d$, has a rate $\sim q^3$,
according to the theory in Ref.\cite{HLR}. This in turn gives a tunneling
current $I\sim \exp{(-{\rm const}/\sqrt{V})}$ for sufficiently small $V$.

The theoretical predictions obtained above from our calculations are in good
overall agreement
with the experiments of Eisenstein {\it et al.}\cite{Jim0,Jim1}. For example,
the peak position obtained in our simple model is quite close to the observed
value of $0.4e^2/\epsilon l_c$. The experimental behavior on the low bias side
of the tunneling peak has also been fit by a formula $I\sim e^{-V_0/V}$,
similar to Eq.(\ref{RR}) above. However, the experimental value $eV_0\approx
0.9e^2/\epsilon l_c$ is much smaller than the value $2\pi e^2/\epsilon l_c$
predicted by our theory. For the range of biases where this behavior is
observed, the wavelength of the density fluctuation contributing significantly
to the relaxation is estimated
to be $\lambda\sim 3l_c$, which is small compared to the inter-layer separation
and the typical
scattering lengths in the system. This implies that inter-layer screening and
impurity scattering should not have a significant effect in modifying the
diffusive behavior of $\chi(q,\omega)$ at the wavelengths and frequencies
involved. Thus it seems that some other factor must be responsible for the
difference between the experimental and theoretical results.
Among the possibilities are that the frequencies probed by the experiment are
still too high for the asymptotic form in Eq.(\ref{AAA}) to be valid; or that
the experimental dependence arises from residual long wavelength fluctuations
in the concentration of the two doping layers, which would tend to smear out
any steep rise in the tunneling characteristic. However, is is also possible
that our theoretical analysis, simply using the RPA value of $\beta$, gives a
poor estimate of the coefficient $\omega_0$ in Eq.(\ref{AAA}).

In summary, we have investigated the properties of the one-electron Green's
function in an interacting
two-dimensional electron system in a strong magnetic field through finite-size
diagonalization and a
low-energy model. We find that the spectral weight is always suppressed at low
energies, with the form of $A(\omega)\sim\exp{(-\omega_0/|\omega|)}$ for the
compressible state at $\nu=1/2$. We have also discussed the implications of our
results on the
electron tunneling into such a system.

\noindent Acknowledgements: The authors are grateful for helpful conversations
with J.~P.~Eisenstein, R.~Ashoori, P.~A.~Lee, and R.~H.~Morf. Work at Harvard
was supported in part by NSF grant DMR-91-15491.

\begin{figure}
\caption{The histograms of the spectral weight $A(\omega)$ of the one-electron
Green's
function at a compressible state obtained for an 8-electron system with
flux number $N_{\phi}=17$. $log_{10}A(\omega)$ is shown in the inset. All
energies are measured from the chemical potential.}
\label{FIG1}
\end{figure}
\begin{figure}
\caption{The histograms of the spectral weight $A(\omega)$ of the one-electron
Green's
function at an incompressible state obtained for an 8-electron system with
flux number $N_{\phi}=21$ corresponding to $\nu=1/3$.}
\label{FIG2}
\end{figure}

\end{document}